\documentclass{elsart}

\usepackage{amssymb}

\begin{document}

\begin{frontmatter}



\title{Nongeneralizability of Tsallis Entropy by means of
Kolmogorov-Nagumo averages under pseudo-additivity} 



\author{Ambedkar Dukkipati,\thanksref{email1}}
\author{M. Narsimha Murty,\corauthref{cor}\thanksref{email2}}
\author{Shalabh Bhatnagar\thanksref{email3}}
\corauth[cor]{corresponding author}
\thanks[email1]{ambedkar@csa.iisc.ernet.in}
\thanks[email2]{mnm@csa.iisc.ernet.in (Tel:+91-80-22932779)}
\thanks[email3]{shalabh@csa.iisc.ernet.in}
\address{Department of Computer Science and Automation,
Indian Institute of Science,\\
Bangalore-560012, India.}

\begin{abstract}
	As additivity is a characteristic property of the classical
        information 
        measure, Shannon entropy, pseudo-additivity of the form $x +_{q}
        y = x + y + (1-q)x y$ is a characteristic property of Tsallis
        entropy. R\'{e}nyi
        in~\cite{Renyi:1960:SomeFundamentalQuestionsOfInformationTheory}
        generalized Shannon entropy by means of 
        Kolmogorov-Nagumo averages, by imposing {\em additivity} as a
        constraint.
        In this paper we show that there exists no generalization for
        Tsallis 
        entropy, by means of Kolmogorov-Nagumo averages, which
        preserves the pseudo-additivity. 
\end{abstract}

\end{frontmatter}


	The starting point of the theory of generalized measures of
	information is due to Alfred
	R{\'{e}}nyi~\cite{Renyi:1960:SomeFundamentalQuestionsOfInformationTheory,Renyi:1961:OnMeasuresOfEntropyAndInformation}, 
	now having an extensive literature. R\'{e}nyi introduced
	generalized information measure, known as $\alpha$-entropy or
	R\'{e}nyi entropy, which is derived by replacing linear averaging
	in Shannon entropy with the generalized averages, in particular
	Kolmogorov-Nagumo averages and by posing the {\em additivity}
	of the information measures. 

	On the other hand, however, Tsallis
	in~\cite{Tsallis:1988:GeneralizationOfBoltzmannGibbsStatistics}
	proposed a non-logarithmic generalization of entropic
	measure, known as $q$-entropy or Tsallis entropy which is
	considered as a useful measure in describing 
	thermostatistical properties of a certain class of physical
	systems that entail long-range interactions, long-term
	memories and multi-fractal systems.

	Tsallis and R{\'{e}}nyi entropy measures are two possible
	different generalization of the Shannon entropy but are not
	generalizations of each other. 

	To understand these generalizations, the so called Hartley
	information 
	measure~\cite{Hartley:1928:TransmissionOfInformation}, of a 
	single stochastic event plays a
	fundamental role. It can be interpreted either as a measure of
	how unexpected the event is, or as a measure of the information
	yielded by the event. In a system with the finite configuration
	space $x = \{x_{k}\}_{k=1}^{n}$, Hartley information measure
	of a single event with probability $p_{k}$ is defined as 
	\begin{equation}
	\label{Equation:Definition_HartlyInformationMeasure}
	H(p_{k}) = \ln \frac{1}{p_{k}}, \quad k=1,\ldots n \enspace. 
	\end{equation}
	Hartley information measure satisfies: (1) H is {\em
	nonnegative}: $H(p_{k})  \geq 0$ (2) H is {\em additive}:
	$H(p_{i}p_{j}) = H(p_{i}) + H(p_{j})$ (3) H is {\em
	normalized}:  $H(\frac{1}{2}) = 1$. These are both necessary
	and 
	sufficient~\cite{AczelDaroczy:1975:OnMeasuresOfInformationAndTheirCharacterization}.
	Hartley information measure can be viewed as a random variable
	and hence we use the notation $H = (H_{1}, \ldots H_{n})$.

	Shannon entropy is defined as an average Hartley
	information:~\cite{Shannon:1948:MathematicalTheoryOfCommunication_BellLabs}  
	\begin{equation}
	\label{Equation:Definition_ShannonEntropy}
	     S (p) = {\langle H \rangle} = \sum_{k=1}^{n} p_{k} H_{k} =
	-\sum_{k=1}^{n} p_{k}\ln p_{k} \enspace. 
	\end{equation}

	The characteristic property of Shannon entropy is additivity,
	i.e., for two independent probability distributions $p$ and
	$r$ we have
	\begin{equation}
	\label{Equation:AdditivityOfShannonEntropy}
	     S(pr) = S(p) + S(r) \enspace,
	\end{equation}
	where $pr$ is the joint distribution of $p$ and $r$. 




	R\'{e}nyi in
	\cite{Renyi:1960:SomeFundamentalQuestionsOfInformationTheory,Renyi:1961:OnMeasuresOfEntropyAndInformation}
	used a well known idea in mathematics  
	that the linear mean, though most widely used, is not the only
	possible way of averaging, but one can define the mean with
	respect to an arbitrary
	function~\cite{HardyLittlewoodPolya:1934:Inequalities,Aczel:1948:OnMeanValues}
	to generalize the Shannon entropy. In the general theory of
	means, a mean of $x= (x_{1}, x_{2}, \ldots, x_{n})$ with
	respect to a probability distribution $p= (p_{1}, p_{2},
	\ldots, p_{n})$ is defined
	as~\cite{HardyLittlewoodPolya:1934:Inequalities}      
	\begin{equation}
	\label{Equation:Definition_KNaverages}
	 {\langle x \rangle}_{\psi} = \psi^{-1} \left[ \sum_{k=1}^{W}
	p_{k} \psi\left(x_{k} \right)    \right] \enspace,
	\end{equation}
	where $\psi$ is continuous and strictly monotonic (increasing
	or decreasing) in which
	case it has an inverse $\psi^{-1}$ which satisfies the same
	conditions; $\psi$ is generally called the Kolmogorov-Nagumo
	function associated with the mean
	(\ref{Equation:Definition_KNaverages})\footnote{A. 
	N. Kolmogorov~\cite{Kolmogorov:1930:SurLaNotionDeLaMoyenne} and
	M. Nagumo~\cite{Nagumo:1930:UberEineKlasseVonMittlewerte} were
	the first to investigate the  
	characteristic properties of general means. They considered
	only the case of equal weights; the generalization to
	arbitrary weights and the characterization of means of form
	(\ref{Equation:Definition_KNaverages}) are due to B. de
	Finetti~\cite{DeFinetti:1931:SulConcettoDiMedia},
	B.
	Jessen~\cite{Jessen:1931:UberDieVerallgemeinerungDesArithmetischenMittels}, 
	T. Kitagawa~\cite{Kitagawa:1934:OnSomeClassOfWeightedMeans},
	J. Acz\'{e}l~\cite{Aczel:1948:OnMeanValues} and many   
	others}. If, in particular, $\psi$ is linear, then 
	(\ref{Equation:Definition_KNaverages}) reduces to the
	expression of linear averaging,
	${\langle x \rangle} = \sum_{k=1}^{n} p_{k} x_{k}$.
	The mean of form (\ref{Equation:Definition_KNaverages}) is
	also referred as quasi-linear mean.

	In the definition of Shannon entropy
	(\ref{Equation:Definition_ShannonEntropy}), if the linear average
	of Hartley information
	is replaced with the generalized average of the form
	(\ref{Equation:Definition_KNaverages}), the information
	corresponding to the probability distribution $p$ with respect
	to KN-function $\psi$ will be
	\begin{equation}
	\label{Equation:Definition_HartleyKNentropy}	
	S_{\psi}(p) = \psi^{-1} \left[\sum_{k=1}^{n} p_{k} \psi \left(
	\ln \frac{1}{p_{k}} \right) \right] = \psi^{-1}
	\left[\sum_{k=1}^{n} p_{k} \psi \left( 
	H_{k} \right) \right] \enspace,
	\end{equation}
	where $H = (H_{1}, \ldots H_{n} )$ is the Hartley information
	measure associated with $p$.

	If we impose the constraint of additivity in $S_{\psi}$, then
	$\psi$  should
	satisfy~\cite{Renyi:1960:SomeFundamentalQuestionsOfInformationTheory}  
	\begin{equation}
	\label{Equation:AdditivityEquationForKNaverages}
	{\langle x + C \rangle}_{\psi} = {\langle x \rangle}_{\psi} +
	C \enspace, 
	\end{equation}
	for any $x=(x_{1},\ldots,x_{n})$ and a constant
	$C$. 

	R\'{e}nyi employed the above formalism to define an
	one-parameter family 
	of measures of information ($\alpha$-entropies)
	\begin{equation}
	\label{Equation:Definition_RenyiEntroy}
	S_{\alpha} = \frac{1}{1-\alpha} \ln \left(\sum_{k=1}^{n}
	p_{k}^{\alpha} \right) \enspace,
	\end{equation}
	where the KN-function $\psi$ is chosen in
	(\ref{Equation:Definition_HartleyKNentropy}) as  
	$\psi(x) = e^{(1-\alpha)x}$, choice motivated by well known
	theorem in the theory of means (Theorem 89,
	\cite{HardyLittlewoodPolya:1934:Inequalities}) that
	(\ref{Equation:AdditivityEquationForKNaverages}) can hold only
	for linear and exponential functions. R\'{e}nyi entropy is an
	one-parameter generalization of Shannon entropy in the sense
	that, the limit $\alpha \rightarrow 1$ in
	(\ref{Equation:Definition_RenyiEntroy}) retrieves Shannon
	entropy.



	On the other hand, Tsallis entropy is given
	by~\cite{Tsallis:1988:GeneralizationOfBoltzmannGibbsStatistics}
	\begin{equation}
	\label{Equation:Definition_TsallisEntropy}
	  S_{q}(p) = \frac{1 - \sum_{k=1}^{n} p_{k}^{q}}{q-1} \enspace,
	\end{equation}
	where $q$ is called nonextensive index ($q$ is positive in
	order to ensure the concavity of $S_{q}$). Tsallis entropy too,
	like R\'{e}nyi entropy, is an one-parameter generalization of
	Shannon entropy in the sense that $q \rightarrow 1$ in
	(\ref{Equation:Definition_TsallisEntropy}) retrieves Shannon
	entropy. The entropic index $q$ characterizes the degree of
	nonextensivity reflected in the pseudo-additivity property
	\begin{equation}
	S_{q}(pr) = S_{q}(p) +_{q} S_{q}(r) = S_{q}(p) + S_{q}(r) +
	(1-q) S_{q}(p) S_{q}(r) \enspace,
	\end{equation}
	where $p$ and $r$ are independent probability distributions.

	Though the derivation of Tsallis entropy, when it was proposed
	in 1988 is slightly different, one can understand this
	generalization using $q$-logarithm
	(see~\ref{Equation:Definition_q-Logorithm}) 
	function: where one would first generalize, logarithm in the
	Hartley information with $q$-logarithm and define $q$-Hartley
	information 
	measure $\widetilde{H}= (\widetilde{H}_{1}, \ldots,
	\widetilde{H}_{n})$ as 
	\begin{equation}
	\label{Equation:Definition_q-HartleyInformationMeasure}
	\widetilde{H}_{k}=\widetilde{H}(p_{k}) = \ln_{q}
	\frac{1}{p_{k}}\enspace, \quad k=1,\ldots n \enspace,
	\end{equation}
	where $q$-logarithm is defined as
	\begin{equation}
	\label{Equation:Definition_q-Logorithm}
	\ln_{q}(x) = \frac{x^{1-q}-1}{1-q} \enspace,
	\end{equation}
	which satisfies pseudo-additivity $\ln_{q}(xy)=\ln_{q}x+_{q}
	\ln_{q}y$ and in the limit $q \to 1$ we have $\ln_{q} \to \ln x$.
	Tsallis entropy (\ref{Equation:Definition_TsallisEntropy})
	defined as the average of q-Hartley information 
	i.e~\cite{TsallisBaldovinCerbinoPierobon:2003:IntroductionToNonextensiveStatisticalMechanics}: 
	\begin{equation}
	\label{Equation:Definition_TsallisEntropy_2}
	S_{q}(p) = {\left\langle \widetilde{H} \right\rangle} =
	{\left\langle \ln_{q} \frac{1}{p_{k}}  \right\rangle} \enspace.
	\end{equation}
	

	Now a natural question arises
	whether one could generalize Tsallis
	entropy in the similar lines of derivation of R\'{e}nyi
	entropy i.e., by replacing linear average in
	(\ref{Equation:Definition_TsallisEntropy_2}) by KN-averages
	under the pseudo-additivity.

	The class of information measures that represent the
	KN-average of $q$-Hartley information measure is written as
	\begin{equation}
	\label{Equation:Definition_q-HartleyKNentropy}
	\widetilde{S}_{\psi} (p) = {\left\langle \ln_{q} \frac{1}{p_{k}}
	\right\rangle}_{\psi} = 
	\psi^{-1} \left[ \sum_{k=1}^{W} p_{k} \psi\left( \ln_{q}
	\frac{1}{p_{k}} \right) \right] = \psi^{-1} \left[
	\sum_{k=1}^{W} p_{k} \psi \left( \widetilde{H}_{k} \right)
	\right] \enspace.
	\end{equation}

	By the pseudo-additivity constraint, $\psi$ should satisfy
	\begin{equation}
	\label{Equation:KNtsallis_PseudoAdditivity_Condition_Form1}
	\widetilde{S}_{\psi}(pr) = \widetilde{S}_{\psi}(p) +_{q}
	\widetilde{S}_{\psi}(r) 
	\end{equation}
	or
	{\setlength\arraycolsep{0pt}
	\begin{eqnarray}
	\label{Equation:KNtsallis_PseudoAdditivity_Condition_Form2}
	\psi^{-1}  && \left[\sum_{i=1}^{n} \sum_{j=1}^{n} p_{i}r_{j} \psi
	\left( \ln_{q} \frac{1}{p_{i}r_{j}}  \right)  \right]  \nonumber\\
	&& = \psi^{-1} \left[\sum_{i=1}^{n} p_{i} \psi \left( \ln_{q}
	\frac{1}{p_{i}}  \right)  \right] +_{q}
	\psi^{-1} \left[\sum_{j=1}^{n} r_{j} \psi \left( \ln_{q}
	\frac{1}{r_{j}}  \right)  \right] \enspace,
	\end{eqnarray}
	where $p$ and $r$ are independent probability distributions
	and $pr$ denotes the joint probability distribution of $p$ and
	$r$.  

	Equivalently, we need
	{\setlength\arraycolsep{0pt}
	\begin{eqnarray}
	\label{Equation:KNtsallis_PseudoAdditivity_Condition_Form3}
	\psi^{-1} && \left[\sum_{i=1}^{n} \sum_{j=1}^{n} p_{i}r_{j} 
	\psi \left( \widetilde{H}_{i}^{p} +_{q} \widetilde{H}_{j}^{r}
	\right)  \right]   \nonumber \\
	 && = \psi^{-1} \left[\sum_{i=1}^{n} p_{i} \psi \left(
	\widetilde{H}_{i}^{p}   \right)  \right] +_{q} 
	\psi^{-1} \left[\sum_{j=1}^{n} r_{j} \psi
	\left(\widetilde{H}_{j}^{r} \right)  \right] \enspace,
	\end{eqnarray}
	where$\widetilde{H}^{p}$ and $\widetilde{H}^{r}$ represents
	the $q$-Hartley information of probability distributions $p$
	and $r$ respectively.

	Note
	that~(\ref{Equation:KNtsallis_PseudoAdditivity_Condition_Form3}) 
	must 
	hold for arbitrary finite discrete probability
	distributions $p_{i}$ and $r_{j}$ and for arbitrary numbers
	$\widetilde{H}_{k}^{p}$ and $\widetilde{H}_{k}^{r}$. If we
	choose $\widetilde{H}_{k}^{r} = J$ independently of 
	$j$  then
	(\ref{Equation:KNtsallis_PseudoAdditivity_Condition_Form3})
	yields that 
	\begin{equation}
	\label{Equation:KNtsallis_PseudoAdditivity_Condition_Form5}
	\psi^{-1} \left[\sum_{k=1}^{n} p_{k}
	\psi \left( \widetilde{H}_{k}^{p} +_{q} J \right)  \right] =
	\psi^{-1} \left[\sum_{k=1}^{n} p_{k} \psi \left(
	\widetilde{H}_{k}^{p} \right) \right] +_{q} J
	\end{equation}

	In general $\psi$ satisfies
	(\ref{Equation:KNtsallis_PseudoAdditivity_Condition_Form5})
	only if $\psi$ satisfies
	\begin{equation}
	\label{Equation:KNtsallis_PseudoAdditivity_GeneralCondition_Form1}
	{\langle x +_{q} C \rangle}_{\psi} = {\langle x
	\rangle}_{\psi} +_{q} C \enspace,
	\end{equation}
	for any $x=(x_{1},\ldots,x_{n})$, which can be rearranged as 
	\begin{equation}
	\label{Equation:KNtsallis_PseudoAdditivity_GeneralCondition_Form2}	
	{\langle x + C + (1-q) x C \rangle}_{\psi} =
	{\langle x \rangle}_{\psi} + C + (1-q) {\langle x
	\rangle}_{\psi} C
	\end{equation}
	or
	\begin{equation}
	\label{Equation:KNtsallis_PseudoAdditivity_GeneralCondition_Form3}
	{\langle (1 + (1-q) C) x + C \rangle}_{\psi} =
	  (1 + (1-q) C) {\langle x \rangle}_{\psi} + C \enspace.
	\end{equation}

	Since  $q$ is independent of other quantities, $\psi$ should
	satisfy the equation of form (By $B =(1 + (1-q) C) $)  
	\begin{equation}
	\label{Equation:KNtsallis_PseudoAdditivity_GeneralCondition_Form4}
	{\langle Bx + C \rangle}_{\psi} = B {\langle x \rangle}_{\psi}
	+ C \enspace.
	\end{equation}

	Finally $\psi$ must satisfy
	\begin{equation}
	\label{Equation:KNtsallis_PseudoAdditivity_GeneralCondition_Sub1}
	{\langle x + C \rangle}_{\psi} = {\langle x \rangle}_{\psi} + C
	\end{equation}
	and
	\begin{equation}
	\label{Equation:KNtsallis_PseudoAdditivity_GeneralCondition_Sub2}
	{\langle Bx \rangle}_{\psi} = B {\langle x \rangle}_{\psi} \enspace,
	\end{equation}
	for any $x=(x_{1}, \ldots, x_{n})$ and for any constants $B$
	and $C$, for $\widetilde{S}_{\psi}$ to preserve the
	pseudo-additivity. 

	From the generalized theory of means
	(\ref{Equation:KNtsallis_PseudoAdditivity_GeneralCondition_Sub1}) 
	is satisfied only when $\psi$ is linear or exponential, but
	the requirement
	(\ref{Equation:KNtsallis_PseudoAdditivity_GeneralCondition_Sub2}) 
	is satisfied only when $\psi$ is linear and it is not
	satisfied when $\psi$ is exponential.
 
	Hence $\psi$ is linear in which case
	(\ref{Equation:Definition_q-HartleyKNentropy}) is nothing but
	Tsallis entropy.

	This establishes the nongeneralizability of Tsallis entropy by
	means of KN-averages under pseudo-additivity. 


\bibliographystyle{elsart-num}
\bibliography{papi}

\end{document}